# Implicit media frames:

# Automated analysis of public debate on artificial sweeteners

Iina Hellsten[1], James Dawson,[2] & Loet Leydesdorff[3]

**Abstract**

The framing of issues in the mass media plays a crucial role in the public understanding of science and technology. This article contributes to research concerned with diachronic analysis of media frames by making an analytical distinction between implicit and explicit media frames, and by introducing an automated method for analysing diachronic changes of implicit frames. In particular, we apply a semantic maps method to a case study on the newspaper debate about artificial sweeteners, published in *The New York Times* (*NYT*) between 1980 and 2006. Our results show that the analysis of semantic changes enables us to filter out the dynamics of implicit frames, and to detect emerging metaphors in public debates. Theoretically, we discuss the relation between implicit frames in public debates and codification of information in scientific discourses, and suggest further avenues for research interested in the automated analysis of frame changes and trends in public debates.

[1] Free University Amsterdam, Athena Institute for Research on Innovation and Communication in Health and Life Sciences, De Boelelaan 1085, 1081 HV Amsterdam, The Netherlands, e-mail: hellsten.iina@gmail.com

[2]University of Amsterdam, Amsterdam School of Communications Research (ASCoR), Kloveniersburgwal 48, 1012CX Amsterdam dawsonanywhere@hotmail.com

[3] University of Amsterdam, Amsterdam School of Communications Research (ASCoR), Kloveniersburgwal 48, 1012CX Amsterdam, e-mail loet@leydesdorff.net

## 1. Introduction

The framing of issues in the mass media plays a crucial role in the public understanding of science and technology (Scheufele, 2007). The dynamics and evolution of frames and sub-debates in public debates have been mainly studied in communication sciences interested in agenda-setting and political communication (e.g., Chilton and Ilyin, 1993; Scheufele, 1999; Nisbet and Mooney, 2007). Digital communications can provide scholars with large amounts of well organized textual data on public debates on science and technology. This chance poses new theoretical and methodological challenges for scholarly research interested in the dynamics of public debates.

Research into the evolution of framing has long roots. As Gamson and Modigliani (1989: 2) formulated: "Public discourse is carried on in many different forums. Rather than a single public discourse, it is more useful to think of a set of discourses that interact in complex ways." In the complex nexus of various competing discourses in the public media, there is a need for tools that provide some coherence to the issues. Frames provide a central organising idea that help to put a particular news item in a wider context. In the mass media, for example, news events are framed in ways that are expected to be familiar to the readers.

Framing essentially involves *selection* and *salience*: "To frame is to *select some aspects of a perceived reality and make them more salient in a communicating text, in such a way as to promote a particular problem definition, causal interpretation, moral evaluation, and/or treatment recommendation* for the item described." (Entman, 1993: 52; italics in the original.) Frames are not static but develop over time with the coverage of a specific topic. The life-cycle of a topic in the mass media has been described by Downs (1972) as an issue-attention-cycle where issues, and the related frames, develop through specific phases. In the analysis of plant biotechnology, Nisbet and Huge (2006) point out that such an issue-attention-cycle fluctuates over time as a series of transitions between technical and dramatic framings, and transitions between political and administrative policy forums.

The public debate on obesity, for example, has changed from an individual's lack of control over eating to a public health problem in reaction to increasing overweight and obesity among the population (Lawrence, 2002). This health issue has turned to the question of whom or what is responsible for causing and curing the emerging epidemic. Is the one who becomes obese also responsible for curing it? The blame and burden in the public debate can be analyzed in terms of systemic *vs*. individualizing frames. An individualizing frame limits the causes of a problem to particular individuals, often the ones suffering from the problem. In other words, an obese person is considered responsible, through personal behavior, for his or her condition. At the opposing pole, systemic frames assign responsibility more broadly to larger social forces such as the food environment, controlled by government and business.

In this sense, the framing of obesity and overweight has fluctuated between medical, political, and economic perspectives. One can expect that the frames manifest themselves in different semantic contexts for word combinations. Can the dynamics of frames be detected via changes in the semantics of the clusters of words in their semantic contexts over time? Recent efforts to automate the scanning of certain public issues and broad topics (Thelwall, Vann, and Fairclough, 2006; Thelwall, Prabowo, and Fairclough, 2006; Thelwall and Hellsten, 2006) have shown that different sets of words may reflect various frames within public debates on specific topics. In this paper we will suggest a way to automate the analysis of the dynamics of implicit frames.

## 2. Explicit and implicit frames

Media frames have been studied intensively in media studies, communication studies, and political communication. Political communication, for instance, has been concerned with the role of framing for media effects (Benford and Snow, 2000; Scheufele, 1999; 2007). In the context of social movements, Gamson and Modigliani (1989) have focused on frames as "interpretative packages." These packages consist of rhetorical devices, such as metaphors, visual images, and symbols. Media framings play a crucial role for the public understanding of science and technology. Research interested in the role of frames in media debates on current issues, has shown that lay people use news media differently from scientists (Nisbet and Mooney, 2007) and for different purposes. Scheufele (2006: 23) notes that "media frames provide audiences with cognitive shortcuts or heuristics for efficiently processing new information, especially for issues that audience members are not very familiar with."

Frames can be implicit or explicit. For example, an author may specifically wish to choose a perspective in order to reduce the complexity of an issue, such as focusing on economic aspects of the issue under discussion. The use of a metaphor may help to narrow the perspective even further. For example, by characterizing genetically modified food as "Frankenstein food" anti-globalists were able to organize resistance against the introduction of genetically modified food without a need to nuance whether in some cases genetic modification would be more detrimental or risky than in others. In such cases, the frames are explicit, and can be investigated with content analysis. Using content analysis, pieces of texts (e.g. statements, paragraphs etc.) are coded to pre-defined categories (e.g. Theme: 1. politics, 2. economics, etc.) (Berelson, 1952; Krippendorf, 1999). Several computer-assisted methods for content analysis have been developed (e.g., Klein, 1991), but they often build on human-induced categories.

The majority of frames and metaphors used in public debates are not explicit. Science, for example, is often approached as “progress,” a step “towards a better future” or “a journey towards specific goals (Hellsten, 2002). However, such frames seem to be so conventionalised that they can hardly be considered explicit. We, therefore, make an analytical distinction between explicit and implicit frames, and argue that this distinction enables further analysis of the development of frames over time.

Implicit frames, we argue, are embedded in latent *dimensions* of the communication, and they are generated because of spurious correlations between word (co-)occurrences in communications. The meanings of words are created in the semantic contexts in which they are used, and such contexts are used to limit the potential scale of polysemous meanings of single words. Words co-occur in text documents at several levels. They may co-occur within sentences and paragraphs in a document, but words can also co-occur across various documents, and sets of documents. The dynamics of co-evolving words in sets of documents that deal with the same topic (e.g., artificial sweeteners) at all these levels may reveal systematic information on latent aspects in communications.

Early efforts to automate the analysis of public debates proposed using words and co-occurrences of words for mapping empirically the translations in the dynamics of science, technology, and society (Callon *et al.*, 1983). In the “sociology of translation” (Callon *et al.*, 1986; Law and Lodge 1984), co-occurrences of words (co-words) have been considered as the carriers of meaning across different domains. Words, however, are ambiguous and languages contain both polysemous and homonymous words. In other words, words are contained within sentences that provide them with meaning (Bar-Hillel, 1955; Hesse, 1980; Leydesdorff, 1995, 1997). The semantic maps approach measures the meanings of words in their contexts, and results in visualizations of word networks. The *relations* between co-occurring words at different levels (within sentences, paragraphs, documents, sets of documents) span a network with an architecture in which words are also *positioned* (Burt, 1982). While word occurrences and co-occurrences can be considered as discrete instantiations of the semantic field, the maps using continuous dimensions can function as a representation of the semantic field itself. These next-order structures can be compared with one another in terms of the degree of codification. In this paper we discuss whether the method allows us to compare implicit frames in public debates over time.

Two artificial sweeteners—saccharin and aspartame, for example—may occur in texts with similar relations with the words "sugar" and "obesity" without necessarily co-occurring themselves in these texts in the same sentences. The similarity in co-occurrence patterns between the two sweeteners can be measured by similarity measures such as the Pearson correlation coefficient, and the components in the networks of words can be visualized by using semantic maps.

In summary, we submit the hypothesis that implicit frames can be indicated using this method and visualization. Unlike explicit frames, implicit frames cannot easily be reconstructed by content analysis. Content analysis focuses on the observable manifestations of communalities contained in the data. The relational data does not contain information about the positions as coordinates on underlying dimensions. These latent dimensions can only be made visible using an algorithmic approach, and by visualising the relative positions of words in relation to each other.[4] We will apply the method of semantic maps to the case study on the artificial sweeteners aspartame and sucralose as reported in

[4] Note that metaphorical meaning is created in the wider textual context than words, such as sentences and paragraphs, and furthermore most metaphors can be expressed in multiple ways, using different words. Therefore, this method is not expected to enable us to detect conventional metaphors in the text, though it may be useful for detecting emerging, novel metaphors.

*The New York Times* between 1980 and 2006 (Dawson, 2007). But let us first provide the wider context of food politics for our case study.

### 3. Politics of food: The case of artificial sweeteners

Food has emerged as a highly charged topic and a much contested field in recent years. Fundamental dilemmas in food production have been exposed relating to risk and control (Lien and Nerlich, 2004). Controversies such as the outbreak of bovine spongiform encephalopathy (BSE) and public debates over genetically modified organisms (GMOs) have placed food at the forefront of political debates.

Food itself has become a political object that has drawn attention to the controversy and conflicts of interests that affect food choice and the structuring of public and media agendas. The notion of risk was introduced to food by the many scandals of the 1990s; for example, the link between BSE in British cows and the Creutzfeld-Jacobs brain disease in humans. The introduction of genetically modified crops led to large-scale protests in Europe, and particularly in the UK. Previously issues like these would have been left up to food-safety authorities and nutrition experts, but nowadays they can be considered as topics of public debate and expert controversy (Lien and Nerlich, 2004).

At the same time, overweight and obesity have been recognised as a public health problem. According to the reportage by Angier, published in 2000 in *The New York Times* obesity has been increasingly been acknowledged as a major public health concern during the last two decades, most notably in the USA where almost two thirds of the adult population suffers from overweight as compared to fifteen percent in the 1980s (Angier, 2000). At the same time, the framing of obesity and overweight has changed from an individual's problem to what Nestle (2002) has called a "toxic food environment.” Obesity has increased since the 1950s hand-in-hand with the growth of processed and sweet-food industries. One reaction to the increasing obesity among the population has been the development of low-calorie foods, such as replacing sugar with artificial sweeteners in a wide range of supermarket products.

The first artificial sweetener, saccharin, was synthesized in 1887 by Remsen and Fahlberg but only widely accepted during World Wars I and II (Bright, 1999). The second artificial sweetener, cyclamate was accepted by the FDA in 1958. In August 1970, concerns over experiments showing that cyclamate induced cancer in laboratory animals led to its ban by the FDA. In January 1970, the FDA decided that diabetics and dieters would be able to use cyclamate on a prescription basis, but by September of that year all foods and drugs containing cyclamate would have to be off the market (anonymous, 1970). However, cyclamate is still used in most countries except the UK and USA, especially in combination with other sweeteners.

More recently, sweeteners such as aspartame, acesulfame-K, sucralose, and neotame have been introduced to the markets. Saccharin, cyclamate, and aspartame are often referred to as “first generation” sweeteners, while sweeteners such as acesulfame-K, sucralose, and neotame have followed and are part of the “second generation” of sweeteners. The synthesis of many new artificial sweeteners in the 1980s and their combination with existing products has led to a broader application of sweeteners as well as much public and media attention. Our focus is on two artificial sweeteners: aspartame (first generation) and sucralose (second generation). Aspartame is known by the trademarks, NutraSweet, Equal, Spoonful, and Candarel; in the EU it has the additive code of E951. The history of aspartame starts in 1965 at the international pharmaceutical company G.D. Searle and Co. (Mazur, 1984). A member of a research group working on an inhibitor for the gastrointestinal secretary hormone, Gastrin, for the purpose of developing an ulcer drug, made an unexpected discovery. A white powder, an intermediate in a chemical reaction, was first tasted by accident and then again intentionally to reveal the intensely sweet taste of aspartylphenylalanine-methyl-ester (aspartame). The tasting event was only reported the following year and it was not until 1969 that the discovery of this artificial sweetener was reported in the *Journal of the American Chemical Society* (Mazur *et al.*, 1969).

In 1970 the FDA banned the artificial sweetener cyclamate, and it was in the same year that testing started to determine the safety of aspartame as a food ingredient. The testing was carried out by G.D. Searle and Company and by their major contractor, Hazleton Laboratories, Inc. On March 5, 1973 G.D. Searle and Company's petition to the FDA for approval to market aspartame as a sweetening agent was published in the Federal Register (FDA, 1973).[5] Eventually more than a hundred studies were submitted to the FDA. The FDA approved aspartame for limited use in 1979. This excluded its use from baked goods, cooking, and carbonated beverages, due in part to the debate growing after initial safety testing that indicated that aspartame may cause cancer in rats (e.g. Ishii, 1981).

A public board of enquiry was convened by the FDA in 1980 in order to discuss the link between aspartame and cancer. The board concluded that aspartame does not cause cancer, but it recommended against approving it at that time due to unanswered questions relating cancer in laboratory rats. The FDA continued to receive complaints of adverse reactions to aspartame use, but could not determine any consistent pattern of symptoms attributed to its use. It was known however that certain people with the genetic disease phenylketonuria and pregnant women with hyperphenylalanaemia have a problem with aspartame. A problem with metabolizing the amino acid, phenylalanine which is one of aspartame's components can lead to brain damage (Ishii, 1981). The FDA ruled, in accepting aspartame in 1981, that all products containing aspartame must warn that the sweetener contains phenylalanine. In 1996, fifteen years after aspartame entered the market, an article published in the *Journal of Neuropathology and Experimental Neurology* by Oltney *et al*. (1996) received a lot of attention from the popular media. The authors—some of whom had been acting against aspartame since its introduction (Smith, 1981)—hypothesized that the increasing human brain cancer rates since 1980 could be linked to aspartame.

[5] Note that the FDA on-line archive starts later than 1973.

Sucralose, in turn, was first approved for use in Canada in 1991 and declared safe by the FDA later in 1998 (FDA, 1998). Over 80 countries now permit the use of sucralose and it can be found in over 4000 products. Sucralose is best known as the trademark, Splenda. In the EU it is has the additive code of E955. Sucralose was discovered in 1975 through a collaborative research project between scientists at Tate and Lyle PLC and researchers at the University of London (Mazur, 1984). Chlorinated sugars were being tested as chemical intermediates in pesticide production, and one was accidentally tasted by one of the researchers revealing its extremely sweet taste.

A young researcher at the Queen Elizabeth College of London, Shashikant Phadnis, misunderstood his supervisor, Leslie Hough, in 1975 when he was told to test a powder and instead tasted the powder. The sweet tasting powder was in fact the novel compound 1',4,6,6'-tetrachloro-1",4,6,6"-tetradeoxygalactosucrose which was synthesised by adding the extremely toxic sulfuryl chloride to a sugar solution. The two researchers saw the potential in the sweet chemical and subsequently worked with Tate and Lyle PLC to refine the formula. The result of over a year's research was a molecule 600 times sweeter than sugar with three chlorine atoms as opposed to the originally tasted compound that had four chlorine atoms. As it turned out the compound was not useful as a pesticide but would prove useful as a food additive.

Artificial sweeteners are loaded with different perspectives that seem to change over time. In this article, we will introduce a method for automated analysis of *implicit* frames in public debates.

This semantic maps method was developed for the analysis of both diachronic and synchronic aspects within sets of documents on scientific topics (Leydesdorff and Hellsten, 2005 and 2006).

## 4. Method: Semantic maps

The analysis of semantic changes in newspaper discourse on artificial sweeteners is based on measuring the meaning of words in their context, and representing the results as networks of co-occurring words after proper normalization (Ahlgren *et al.*, 2003; Leydesdorff and Vaughan, 2006). The semantic maps method builds upon the following procedure. The similarity in occurrence patterns of words between the words can be measured by similarity measures such as the Pearson correlation coefficient. The similarity matrix can be further analyzed using factor analysis or multi-dimensional scaling in order to obtain a visualization of the network in terms of its main dimensions (Kruskal and Wish, 1978; Schiffman *et al.*, 1981).

Based on similar principles as multidimensional scaling—that is, the reduction of stress in the projection onto a two-dimensional map—visualization programs of social network analysis allow us to visualize not only the positions of words as nodes in the network, but also the strength of their relations as links. In a seminal article, Kamada and Kawai (1989) reformulated the problem of achieving graph-theoretical target distances in terms of energy optimization. This algorithm uses a gradient descent method to iteratively minimize the stress in network visualizations.

Because one can expect the distribution of words to be heavily skewed (Ijiri and Simon, 1977; Chen, 1985), the Pearson correlation coefficient (which assumes a normal distribution) is not the appriopriate similarity criterion (Ahlgren *et al.*, 2003). Salton and McGill (1983) proposed to use the cosine instead. The cosine is similar to the Pearson correlation coefficient with the difference that one does not normalize using the arithmetic mean of the distribution, but uses the geometrical mean (Jones and Furnas, 1987). Therefore, this is also called the vector-space model (Salton and McGill, 1983). The cosine is formulated as follows:

$$\text{Cosine}(x,y) = \frac{\sum_{i=1}^{n} x_i y_i}{\sqrt{\sum_{i=1}^{n} x_i^2}\sqrt{\sum_{i=1}^{n} y_i^2}} = \frac{\sum_{i=1}^{n} x_i y_i}{\sqrt{(\sum_{i=1}^{n} x_i^2)*(\sum_{i=1}^{n} y_i^2)}} \quad (1)$$

where $x_i$ and $y_i$ refer to the score of the $i^{th}$ row (e.g., document) in column $x$ or $y$ (e.g., different words).

Using the cosine matrix as input to the visualization, one visualizes the vector space. The vector space has a topology different from the relational space since it represents coordinates. Distances in it are based on similarity in the distributions of words in documents and not on the relations among words (Leydesdorff and Vaughan, 2006).

This so called semantic maps method has previously been applied (with success) to the analysis of codification in scientific texts on nanotechnology (Lucio-Arias and Leydesdorff, 2007), comparison between several discourses on one topic, stem cell research (Leydesdorff and Hellsten, 2005), and tracing the development of one debate over time as well as comparing one single text to a set of texts (Leydesdorff and Hellsten, 2006). In this paper, we test whether the method is able to automatically trace emerging implicit frames in public debates on science and technology.

The comparison across various discourses in the stem-cell debate (Leydesdorff and Hellsten, 2005) showed that newspaper articles codified this debate to a lower extent than social-scientific debates, whereas title-words in the natural and life sciences provided mainly variation, i.e., the level of codification was highest in social scientific articles, and lowest in the natural and life sciences, as measured by the title words of articles indexed in the ISI's Web of Science, social sciences and science databases. In the natural and life sciences, codification is achieved by highly-disciplined citation behavior (Leydesdorff, 1989; 1995).

Leydesdorff and Hellsten (2006) added the distinction between restricted and elaborate discourses by comparing the results the semantic maps method provided for single texts and sets of texts (Bernstein, 1971). A single author cannot be expected to change the meanings of words as frequently as these meaning can be changed within discussions. In the case of a single author, one can expect a more densely packed network of words and hence a more stable frame. In the analysis, in case of such a restricted discourse, one may need to set the value of the threshold used for the cosine higher than in the case of a less dense network.[6]

In this study, we apply this method, that seems to produce promising results, to newspaper articles on the artificial sweeteners “aspartame” and “sucralose” in *The New York Times* between 1980 and 2006. We conducted the analysis with three, partly overlapping data sets, each of which was collected with different search terms. All three data sets contain newspaper items published in *The New York Times* in between 1980 and 2006, and in each case snapshots of the most frequent periods of debate were selected for further analysis. First, we collected the data using the general term “artificial sweetener;” second, items using the term “aspartame” were downloaded; and third, the term “sucralose” was used for the data retrieval.

Each newspaper article containing one of these words in the title was downloaded and saved as a separate text file. Thereafter, capital letters were transferred into small letters, and the plural s was removed (e.g., the words “car” and “cars” were stemmed into the single word, “car”). The stop word list from the U.S. Patent and Trade Office (USPTO) was used throughout the analysis.[7] Word frequency lists were created of each document using the concordance program Text STAT.[8] The texts and words were inputted into dedicated software programs freely available at http://www.leydesdorff.net/indicators/index.htm. These programs produce the asymmetrical matrices word/document occurrence matrices, the co-word matrices, and the cosine-normalized similarity matrices among the words based on the asymmetrical word/document matrices (Leydesdorff and Vaughan, 2006).

---

6 Unlike the Pearson correlation coefficient which varies from –1 to +1, the cosine varies from zero to one and thus one has to choose a threshold level. Without a threshold most nodes would be connected to each other with a cosine > 0. (Egghe and Leydesdorff, in preparation).

7 The list is available at http://www.uspto.gov/patft/help/stopword.htm

The output files of the programs can be imported in the visualization program Pajek[9] in order to visualize the results in the form of semantic maps based on similarities among word distributions using the cosine for the normalization. The mean of the cosine of the lower triangle of the matrix was used as a cut-off level for word frequencies in order to optimize the visualizations. Approximately 100 words at the maximum were included in the analysis. Although including many more words is possible, the reading of the semantics maps in this case becomes problematic. Generating the visualization, a procedure that is included in the Pajek program, involves iteratively repositioning nodes to minimize the total "energy" of the spring system using a steepest descent procedure. The size of the nodes is proportional to the logarithm of the frequency of the word occurrences in the matrix. A disadvantage of the approach is that unconnected nodes may appear randomly positioned in the map. Therefore, unconnected nodes were removed from the visualizations.

For guiding the interpretation of the semantic maps that are often very rich in information, we decided to focus in particular on a set of words that were common to all the data sets. For this purpose, sets of word frequency lists were compared to each other. The five most commonly occurring words were, not surprisingly, "product," "sweetener," "food," "sugar" and "diet." Of these five words, only "product" occurred in all word frequency lists, while "sweetener" appeared but in one, “food” in all except two and “sugar” and “diet” but in three of the lists. Even though these five words do not all appear in all texts, we will use them as a focus for the interpretation of our results. There is need for further testing with frequently and less frequently used words as a focus of interpretations.

---

8 TextSTAT is a freeware program from the Free University of Berlin available at http://www.niederlandistik.fu-berlin.de/textstat/software-en.html

9 Pajek is available at http://vlado.fmf.uni-lj.si/pub/networks/pajek/ as freeware.

## 5. Results

Our analysis starts from the semantics of artificial sweeteners in general and proceeds into semantic changes in one of the first generation sweeteners, aspartame, and one of the second generation sweeteners, sucralose.

### *5.1 Artificial Sweeteners*

*The New York Times* published 54 newspaper items using the term “artificial sweetener” during the period between 1980 and 2006. The number of published articles reached a peak in between 1984 and 1986 (16 published items) and again from 2004 to 2006 with eight new news items. Further analysis of the semantics of this set of newspaper items that mentioned the term “artificial sweetener” is focused on the peaks in frequency of published items around the year 1985 and 2005. We included all items for the 3-year periods from 1984 to 1986, and from 2004 to 2006. The full body of each article was saved in plain text format. Using a word list of 60 words occurring more than ten times in the 16 articles from 1984-1986, the semantic map shown in figure 1 was generated. The mean of cosine in lower triangle when zeros are not included was used as a threshold for the visualization in Pajek; this threshold value was cosine ≥ 0.46 in this case.[10]

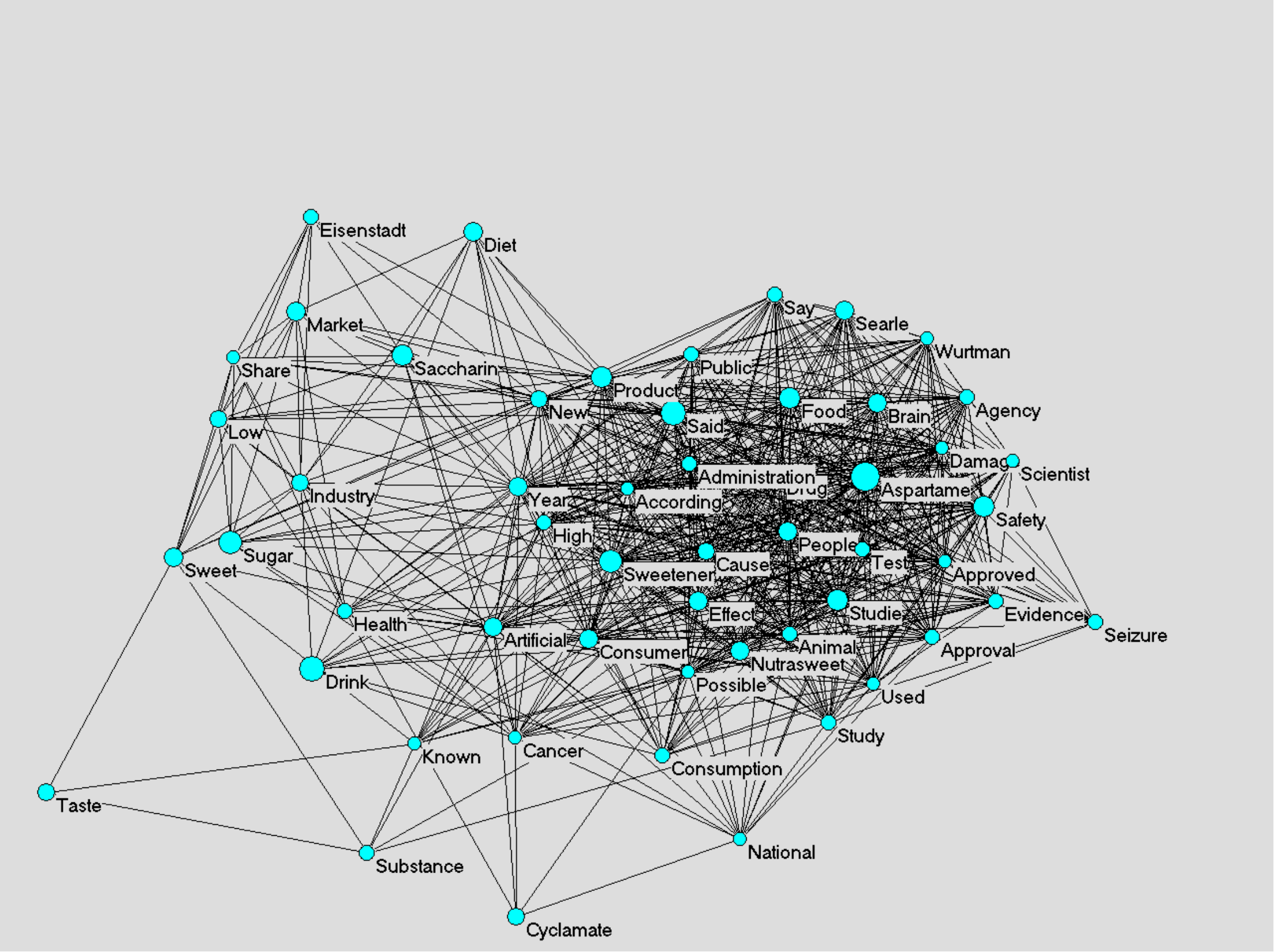

Figure 1: Semantic map of 60 words used more than ten times in 16 articles containing the string “artificial sweetener” in the *NYT* between 1984 and 1986 (cosine ≥ 0.46).

Analysis of the visualization of the articles published between 1984 and 1986 provides a good idea of what was discussed at this time in the *NYT*. Interestingly, all three of the first generation sweeteners are present with “saccharine” most closely associated with “diet” and “market” in the map, and “cyclamate” as the most closely associated with “cancer.” Despite the ban of cyclamate by the FDA more than ten years earlier, it was present as a negative dimension of the framing in the journalistic discourse.

[10] The mean value is often used as a cut-off point in social network analysis (Wasserman and Faust, 1994, at pp. 406 ff.). For example, in UCINet one can choose “MEAN” as a distinguishing value for making a matrix binary. However, the mean including zeros (in this case, 0.44) is usually too low for a meaningful representation because of the skewness in the (Lotka-) distributions.

The word “aspartame” is in the centre of a tightly grouped cluster of administrative words such as “food,” “administration,” “approved,” and “study” that are relatively close to the search term “artificial sweetener.” This shows that there was still much discussion over the safety of aspartame as words like “seizure,” “brain,” and “damage” are closely associated with it.. In this earlier snapshot the word “diet” is found on the edge of a weak secondary cluster with many connections. Connections to words like “product,” “market,” and “drink,” and the separation from the main cluster of the map give the impression that “diet” is not yet central to the discourse around artificial sweeteners. Using a list of the 48 words used more than ten times from the eight texts in the *NYT* between 2004 and 2006, the semantic map shown in figure 2 was generated.

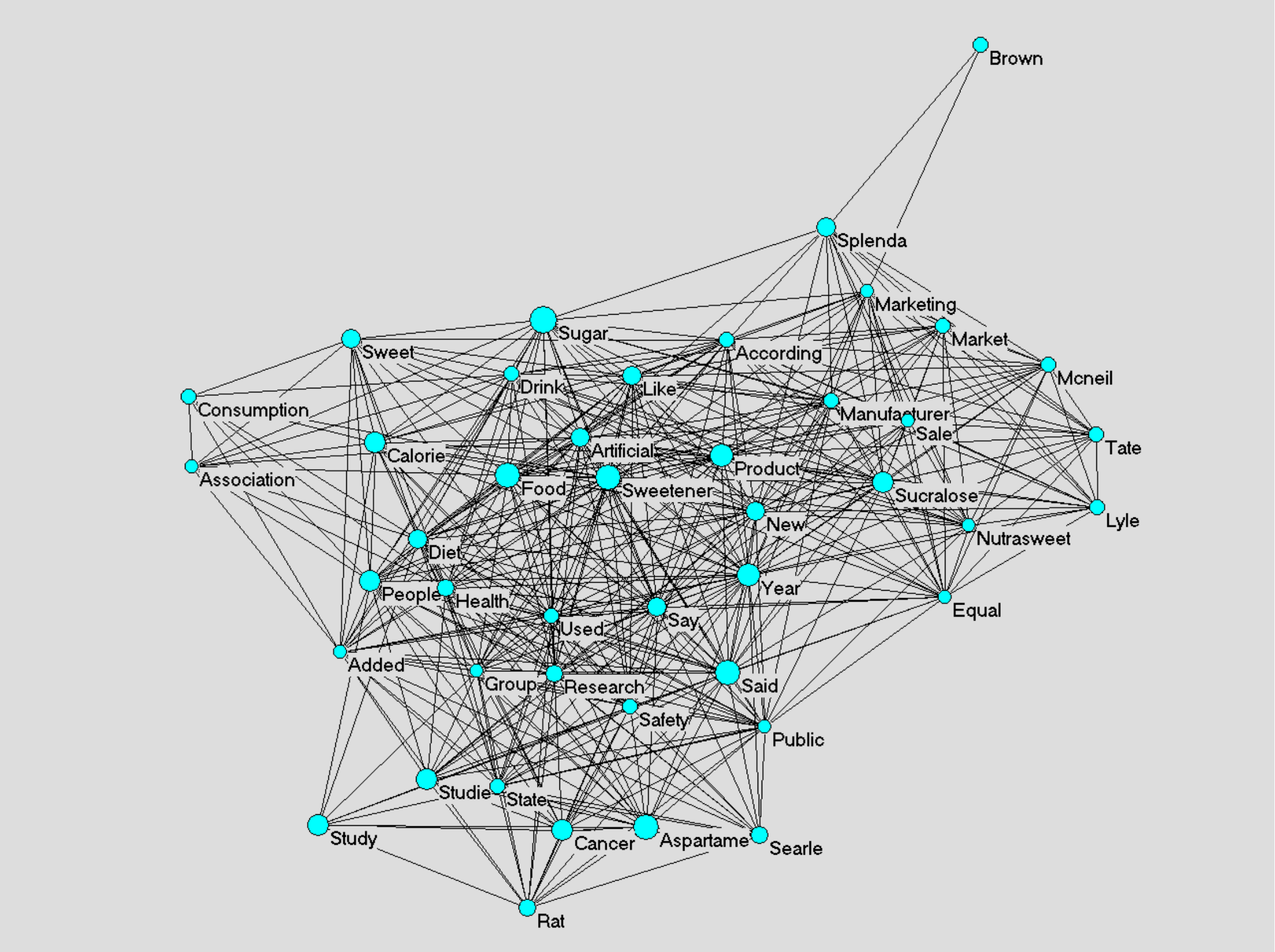

Figure 2: Semantic map of 48 words used more than ten times in eight articles containing the string “artificial sweetener” in the *NYT* between 2004 and 2006 (cosine ≥ 0.51).

In this map, the first-generation sweeteners cyclamate and saccharine are no longer represented. Acesulfame-K and the other second-generation sweeteners are also conspicuous in their absence. “Sucralose” is now most closely associated with the search term “artificial sweetener.” This most recently approved sweetener is more prominently related to the names of the two companies responsible for its marketing and manufacture, *viz.* “McNeil” and “Tate and Lyle”, respectively.

“Aspartame” is still closely associated with words like “cancer” and “study” in a loose cluster slightly separated from the main cluster of the map. Interestingly the word “Searle” is close to aspartame even though as the company responsible for the initial discovery and marketing of aspartame, G.D. Searle and Co., was acquired by Monsanto in 1985. “Food,” “diet,” “health,” and “sweetener” are now central in the main cluster of this network.

### *5.2 Aspartame*

Despite the suggested link between aspartame and brain cancer, the FDA approved aspartame for general use in 1996 including all baked goods and beverages. The only remaining limitation is a maximum of 0.5% aspartame content in baked goods, but this still allows aspartame to be used as the sole sweetener in any baking application. Aspartame is now used in more than 5000 products and consumed by over 200 million people around the world. In this section we focus on the representations of aspartame in *The New York Times* from 1980 to 2006, and thereafter the changing semantics of newspaper texts during this same period of time.

As expected, the frequency of published newspaper items on aspartame peaks in between 1983 and 1986. This is likely due to the acceptance of aspartame and its entrance to the consumer market at the time. Two events significant to aspartame and artificial sweetener occurred in 1985. Firstly, Monsanto acquired G.D. Searle and Co., the inventors of aspartame, and, secondly, the National Institute of Health determined for the first time that obesity was a major threat to public health (Lawrence, 2004). After this signal, however, the number of articles published in the *NYT* about aspartame steadily decreased.

To complement the visualizations generated for "artificial sweetener," we analyzed the word occurrences in the years around 1981, that is, before the acceptance of aspartame by the FDA, and for the period of 2004 to 2006. A semantic map was generated using a word list of 55 words occurring more than five times in the six articles published between 1980 and 1982.  (Figure 3).

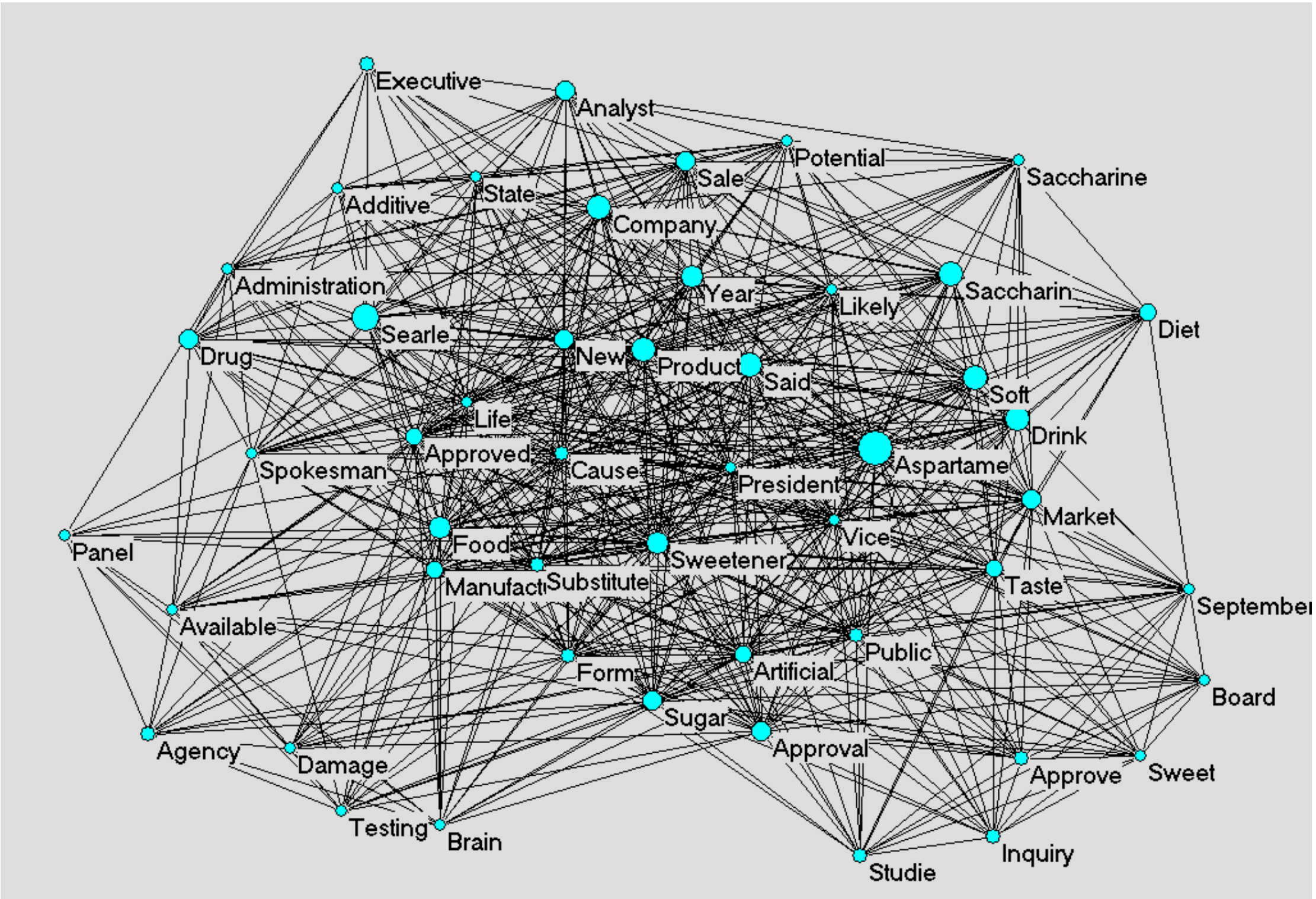

**Figure 3**: Semantic map of 55 words used more than five times in six articles containing the word "aspartame" in the *NYT* between 1980 and 1982 (cosine $\geq 0.54$).

Around the time of the approval of aspartame in 1981, the core cluster of the map in figure 3 is made up of administrative words, such as "spokesman," "drug," "approved," and "additive." The only other first-generation sweetener still available at this time, saccharine is quite prominent and appears relatively close to "aspartame" in the network. Interestingly, a weak cluster of words like "brain" and "damage" appears on the bottom left of the network but the word "cancer" is not present using this threshold for the visualization. The words "food," "product," "sweetener," and "sugar" are all present in the major, tight cluster of this map, while "diet" is somewhat removed. However, "diet" is again well connected with words such as "soft" and "drink."

Searching the *NYT* online database between 2004 and 2006 for "aspartame NOT sucralose" yielded eight texts. A word list of 45 words used more than six times was generated, after removing stopwords, from the eight texts published between 2004 and 2006 (Figure 4).

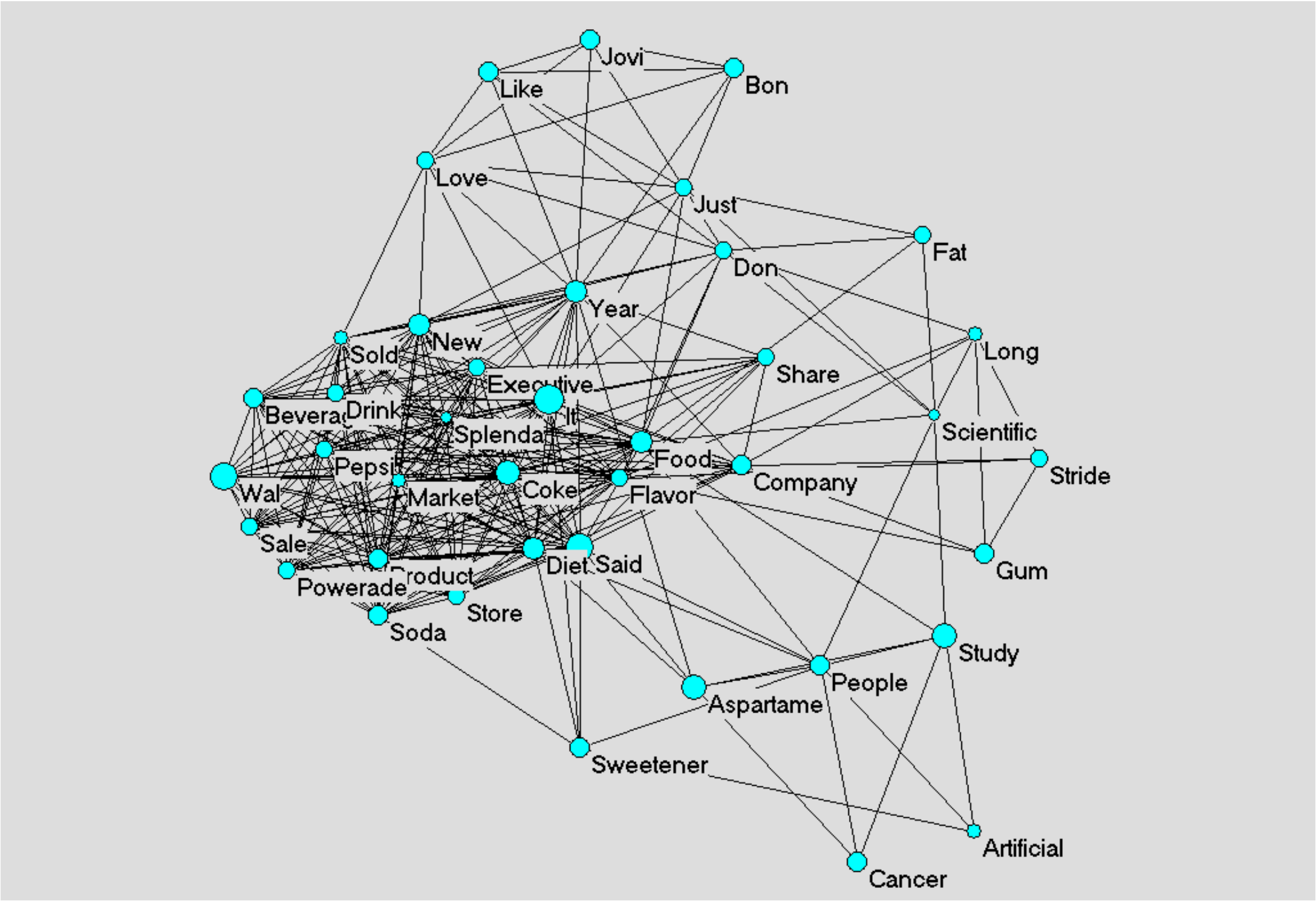

**Figure 4**: Semantic map of 45 words used more than once in ten articles containing the word "aspartame" in the *NYT* between 2004 and 2006 (cosine ≥ 0.482).

This more recent map (Figure 4) shows a dense cluster around the use of "splenda" in beverages such as Coke, Pepsi, and Soda as diet products. There are several more weakly connected words and clusters as well. Noteworthy, there is a weak cluster of the words "Bon" and "Jovi" at the top of the figure. At first, it seems like a mistake to find "Bon Jovi" in newspaper articles related to "aspartame," but further inspection of the data showed that indeed aspartame appears in an article in which Bon Jovi is mentioned several times. In that article the term "aspartame-infused" is used to describe some of Bon Jovi's songs. This term in the article, along with the other words present in the network of figure 4 represents a meaning of aspartame codified differently from the two decades earlier. The major cluster of the network includes the word "Splenda." Business-related words and "cancer" are now closely related to aspartame. At this time, “diet” is much more centrally located near “product” and “market.”

### *5.3 Sucralose*

Sucralose was accepted by the FDA in 1998, and is in use in thousands of products worldwide. Searching the *NYT* online database for texts containing the word “sucralose” was carried out in much the same way as the search for “aspartame” above. In this case the search term used was “sucralose NOT aspartame.” Excluding “aspartame” from the search reduced the number of results from 32 to 15.

Only one text was published in the *NYT* between 1997 and 1999 containing the word “sucralose.” This article was published in the year of the FDA approval, 1998. As a single document, unlike document *sets*, can be expected to provide a restricted discourse with well organized and tightly connected word usage, we only draw a semantic map on a later point of time. The most articles per year mentioning “sucralose” occurred in 2005 with five articles, this compared to the 37 articles mentioning aspartame in 1983. In between 2004 and 2006, five texts were published. These texts were used to create a frequency list of 47 words, used more than five times (excluding stop words) (Figure 5).

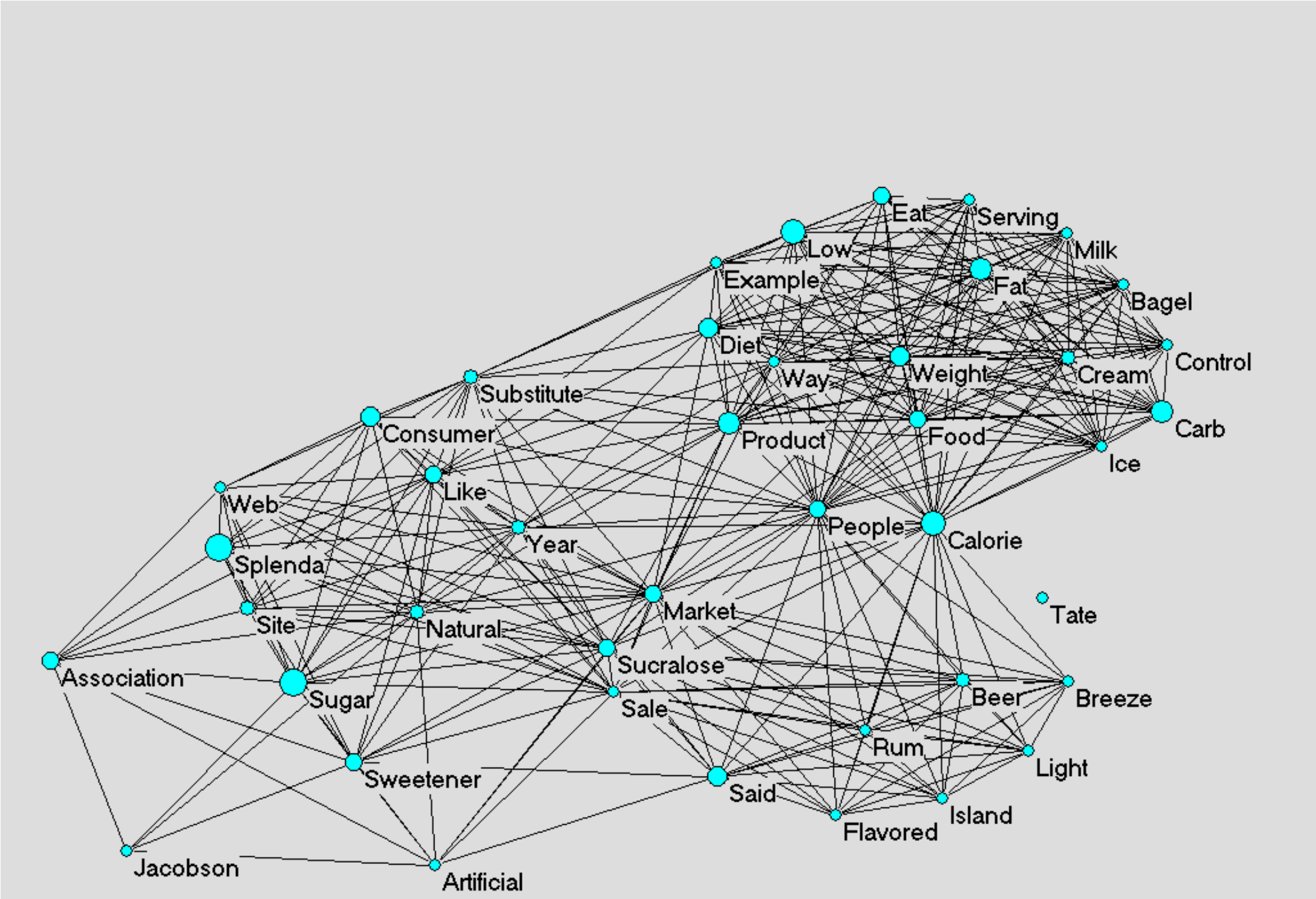

**Figure 5**: Semantic map of 47 words used more than five times in five articles containing "sucralose" in the *NYT* between 2004 and 2006 (cosine ≥ 0.54).

Interestingly in this map, "sugar" is even more centrally located than "sucralose" itself. This is especially relevant in the light of recent "splenda" marketing campaigns which emphasize the close chemical relationship between sucrose and sucralose. "Splenda" has more presence than "sucralose," but remains still closely associated with "sugar."

**6. Discussion**

In this article we have discussed the applicability of the semantic maps method for automatically detecting and tracing the development of implicit frames in public debate on artificial sweeteners over a period of twenty years in one newspaper, *The New York Times*. We focused on one newspaper to test the validity of the method without the disturbance of differences among different newspapers or other types of mass media. There is need for further research that analyses a wider set of newspaper items, or data from different types of mass media, such as newspapers, magazines and on-line news services in order to further validate the proposed method.

In summary, close inspection of the semantic maps generated from the *NYT* texts reveals an abundance of administrative words in several of the sub-debates. The organisation responsible for regulating the use of technology such as artificial sweetener, the FDA, is almost always mentioned alongside the sweetener itself. The semantic maps of the texts around 1985 indicate that there is still much discussion about the safety aspects of the sweetener technology alongside the administrative repertoire. The semantic maps of later texts show less discussion of the health risks of the sweeteners amongst the administrative words, and more discussion of the health benefits. Diet, food, and artificial sweetener become more closely associated over time from 1985 to 2005. It is also significant that only one article was published about sucralose in a 3-year period around the year (1998) in which it was accepted by the FDA. In comparison, six articles published in connection with aspartame in a 3-year period surrounding its acceptance by the FDA (in 1981). The acceptance of a new sweetening technology seems to be less newsworthy in 1998 than it was in 1981.

In particular, there are three points for further discussion. First, in all the three, partly overlapping, data sets, the word “diet” moves from the outskirts of the semantic map into the central, dense clusters over time. This can perhaps be considered as an indication of frame change over time from the individual to the more systematic frame of approaching and discussing obesity and overweight. There is need for further research on this possible frame change within the public debate on obesity and overweight. Our focus was narrowed to artificial sweeteners.

Second, the appearance of the metaphor “aspartame-infused” in 2005 aspartame texts from the *NYT* is perhaps the clearest indication of this change in codification over time. If the author had chosen instead to use “sucralose-infused,” this might have meant something to significantly fewer people. The appearance of metaphoric use of an artificial sweetener as well as the reduction in *NYT* publishing activity indicate increased public acceptance of artificial sweeteners. The metaphorical use of aspartame around the year 2005 builds upon the expectation that aspartame is well know to most people, and that some of its associations, in particular as sickly sweet and artificial, are shared by a wide group of people. Without a wide acceptance of the popular associations of aspartame, the metaphor of Bon Jovi’s music and this particular sweetener would be non-understandable.

Third, using the five most frequently used words from all of the frequency lists of all of the data sets provided a restricted analytical instrument. The frequency lists from each data set already had stopwords removed and represented only words used more than a given number of times. In each case this given number was specified but was in fact different. For the purposes of the study, however, using the words “product,” “sweetener,” “food”, “sugar,” and “diet” as they were the most frequent among the most frequently used, provided a heuristic. There is need for further research into the use of selected keywords in guiding the interpretation of results on large sets of texts that otherwise may remain beyond systematic analysis.

**7. Conclusion**

In conclusion, our results show that the method seems promising for capturing changes in the implicit frames in public debates. First, the analysis of co-occurrences between words and the distribution of these words in a set of texts on a pre-specified topic (such as artificial sweeteners), and the semantics of words can inform us about the (implicit) frames in media debates. Furthermore, implicit frames in newspaper texts seem to function similarly to codification in scientific texts. Both framing (in the media) and codification (in scientific texts) provide coherence to otherwise seemingly unstructured flows of information. In scientific discourse meaning is further codified in disciplinary jargons, while in newspapers natural language remains a first-order codifier (Leydesdorff and Hellsten, 2005). The frames can therefore be expected to remain more fluid than scientific codifications. This finding has potentially wide-spread theoretical implications for research interested in meaning and information processing in communication systems.

Second, the method was able to detect an emerging metaphor, Bon Jovi's music as aspartame-infused, within this debate. The appearance of the metaphoric use of aspartame connected with the music of Bon Jovi raises an interesting question. The phrase "aspartame-infused" is used by the author to describe Bon Jovi's romantic rock ballads as excessively romantic or unnatural. There are many metaphors relating to sweetness that are well known and occur relatively often especially in the discourse about artificial sweeteners. In a business sense it is not unusual to read of a sweetener that has soured while health concerns are often referred to as bitter-sweet. The method was able to detect this highly unusual, emerging metaphor as an implicit frame in the set of documents. This is in accordance with a previous finding that metaphors may function as anti-codifiers (Hellsten and Leydesdorff, 2005).

Third, the method can be automated, and therefore made suitable for the analysis of large amounts of texts on public debates on science and technology, presently available in digital format in various databases. In this sense, the method is a contribution to the problem of dealing with increasing amounts of data in research interested in public communication of science and technology. The analysis in this paper was still based on comparing snapshots for different moments of time, but recently, programs have been developed that will allow one to animate a series of cosine matrices (as representations of the vector space) over time (Leydesdorff and Schank, 2008; Leydesdorff *et al*., in preparation).

Theoretically, our results point to next-order meaning-processing in communications that emerges on top of the information-processing (Leydesdorff, 2001). However, the construction of meaning (via either framing or codification) can be expected to feed back on the information processing in evolving systems (Luhmann, 1984; 1995). The construction of communication systems works from the bottom up while control (framing / codification) tends to function top-down. The complex dynamics of framing and codification are structured in latent dimensions of communications which interact and co-evolve with each other. There is urgent need for further research into automated analysis of public debates, the dynamics of implicit and explicit framing, and the relations between framing and codification as potentially parallel processes of the cultural construction of meaning in society.